**Electrical Conductivity of Superionic Hydrous SiO$_2$ and the Origin of Lower-mantle High Conductivity Anomalies Beneath Subduction Zones**

**Mako Inada[1]\*, Yoshiyuki Okuda[1]\*, Kenta Oka[1], Hideharu Kuwahara[2], Steeve Gréaux[2], and Kei Hirose[1,3]**

[1] Department of Earth and Planetary Science, The University of Tokyo, Hongo, Tokyo 113-8654, Japan

[2] Geodynamics Research Center, Ehime University, Matsuyama, Ehime 790-8577, Japan

[3] Earth-Life Science Institute, Institute of Science Tokyo, Meguro, Tokyo 152-8550, Japan

**\***Corresponding author: Mako Inada (maccolo-0515@g.ecc.u-tokyo.ac.jp), Yoshiyuki Okuda (okuda.y@eps.s.u-tokyo.ac.jp)

**Key Points:**

- We measured the electrical conductivity of hydrous Al-bearing SiO$_2$ up to 82 GPa and 2610 K.
- Conductivity sharply increased at high temperature, attributed to the superionic transition.
- Superionic silica in subducted MORBs may cause high electrical conductivity anomalies.

**Abstract**

Electrical conductivity (EC) is one of the important physical properties of minerals and rocks that can be used to characterize the composition and structure of the deep interior of the Earth. Theoretical studies have predicted that the $CaCl_2$-type hydrous Al-bearing $SiO_2$ phase, present in subducted crustal materials, becomes superionic —meaning that protons are no longer bonded to a specific oxygen atom but instead become mobile within the $SiO_2$ lattice— under high-pressure and high-temperature conditions corresponding to the lower mantle. The enhancement of the EC upon such superionic transition has not been experimentally verified yet. Here, we measured the EC of Al-bearing $SiO_2$ containing 1750 ppm $H_2O$ at pressures up to 82 GPa and temperatures up to 2610 K by employing a recently developed technique designed for measuring transparent materials. Results demonstrate a sudden increase in EC to approximately 10 S/m at temperatures of 1100–2200 K, depending on pressure, which is several to ten times higher than that of the surrounding shallow to middle part of the lower mantle, which is attributed to a transition to the superionic state. If hydrous $SiO_2$ is substantially weaker than other coexisting phases and thus forms an interconnected film in subducted MORB crust, the EC of the bulk MORB materials is significantly enhanced by superionic $SiO_2$ in the lower mantle up to ~1800 km depth, which may explain the high EC anomalies observed at subduction zones underneath northeastern China. The observed EC anomalies can be matched by the EC of subducted MORB materials containing Al-bearing $SiO_2$ with a water content of approximately 0.2 wt%, providing insights into the deep $H_2O$ circulation and distribution in the Earth's mantle.

**Plain Language Summary**

The EC of minerals, when combined with geophysical observations, provides valuable insights into materials and processes that occur deep within the Earth. In particular, the high ECs of water-bearing minerals help us map deep circulations and distributions of water in the mantle. In subduction zones beneath Northeast Asia, there are anomalous areas exhibiting notably higher EC compared to the surroundings. In our laboratory study, the EC of water-containing $SiO_2$ was measured under high-pressure and temperature (*P-T*) conditions. The results show a sudden increase in EC with increasing temperature, validating a theoretical prediction of a transition to the superionic state; protons move freely within the $SiO_2$ crystal, which greatly enhances its ability to conduct electricity. The observed *P-T* range of this superionic state suggests that water-containing $SiO_2$ present in subducted crustal materials exhibits exceptionally high EC in the lower mantle, extending to depths of approximately 1800 km. The high EC anomalies observed beneath subduction zones in Northeast Asia may be explained by the presence of superionic water-containing $SiO_2$.

## 1. Introduction

Superionicity describes a process of electrical conduction in materials in which highly mobile cations or anions exhibit diffusive behaviors throughout the lattice in the presence of an electric field. Such a unique property has been found in mantle minerals of $FeOOH_x$ (Hou et al., 2021), where protons are diffusive, and K-hollandite (liebermannite) (Manthilake et al., 2020), where potassium ions show similar behavior. In these minerals, respective cations are no longer bound to specific anions but instead move fluidly within the solid anion framework. It has also been predicted for $H_2O$ ice at high *P-T* (Cavazzoni et al., 1999) and experimentally confirmed through optical conductivity measurements (Millot et al., 2018; Prakapenka et al., 2021). The phase diagram of $H_2O$ has been extensively investigated (Gleason et al., 2022; Husband et al., 2024; Iwamatsu et al., 2024; Millot et al., 2019; Prakapenka et al., 2021; Weck et al., 2022). With its superionicity, ice has enhanced EC, where protons diffuse rapidly through the oxygen sublattice exhibiting fast ionic conduction (Cavazzoni et al., 1999; Sun et al., 2020). Earlier molecular dynamics simulations predicted such large proton mobility not only in $H_2O$, but also in hydrous Al-free (Li et al., 2023) and Al-bearing dense $SiO_2$ (Umemoto et al., 2016) at 20–60 GPa and ~1000–1500 K corresponding to conditions for subducting former oceanic lithosphere (slabs) represented by the cold lower-mantle geotherm. The Al-bearing $SiO_2$ is a major phase in subducting crustal materials, constituting 15–30% of mid-oceanic ridge basalt (MORB) compositions (Hirose et al., 2005; Ono et al., 2001; Ricolleau et al., 2010) and 20–75% of continental crust compositions (Irifune et al., 1994; Komabayashi et al., 2009) at lower-mantle depths. Hydrous $SiO_2$ is the only major phase included in the slabs that has been predicted to have superionicity. It has been argued that once subducted MORB crusts are dehydrated, typically at depths shallower than at least ~300 km (Okamoto & Maruyama, 1999), they are rehydrated when hydrous phases such as superhydrous phase B and phase D decompose in the underlying peridotite layer of the slab around 700 km depth (Ohtani, 2020; Walter, 2021). The reactions of MORB with $H_2O$ fluid (a hydrous melt) at that depth form hydrous Al-bearing $SiO_2$ stishovite (or the $CaCl_2$-type phase) (Amulele et al., 2021; Ishii et al., 2022), which is a main carrier of $H_2O$ in slabs subducted in the lower mantle.

High EC anomalies have been detected at depths of ~700 km beneath subduction zones in northeastern (NE) China and of ~900–1200 km beneath the Japan Sea (Kelbert et al., 2009; Kuvshinov, 2012). Since none of the mineral phases, including hydrous phases, were known to have high ECs comparable to the observed anomalies in this region (Guo & Yoshino, 2013), these anomalies have traditionally been attributed to partial melting in the uppermost lower mantle induced by $H_2O$ released from dehydrating slabs (Koyama et al., 2006; Schmandt et al., 2014). However, partial melting cannot explain the observed high EC anomaly since the EC of silicate melt is not high enough when considering its limited volume fraction (Okuda et al., 2024; Zhang et al., 2021). On the other hand, seismological studies have suggested that segregated MORB materials are piled in these regions (Feng et al., 2021; Li & Yuen, 2014; Niu, 2014). The presence of superionic hydrous Al-bearing $SiO_2$ could therefore potentially explain such high EC anomalies. Despite its importance, its superionicity has yet to be experimentally demonstrated because the EC measurements are still challenging at high *P-T* conditions of the lower mantle.

Here, we measured the EC of Al-bearing $SiO_2$ containing 1750 ppm $H_2O$ at 41–82 GPa and 660–2610 K based on laser-heated diamond-anvil cell (DAC) techniques. We employed a method recently developed for measuring the electrical resistance of a material that is transparent to a near-infrared laser beam used for heating samples in a DAC (Okuda et al., 2022). The obtained EC exhibited a sudden increase at the temperature range of 1100–2200 K depending on pressure, likely caused by a transition to the superionic state. The high EC of superionic hydrous silica may explain the observed high EC anomalies beneath the subduction zones in NE China.

## 2. Experimental Methods

2.1 Sample preparation

A hydrous Al-bearing $SiO_2$ sample was synthesized from an $SiO_2$ glass + 7.4 wt% $Al(OH)_3$ (2.6 wt% $H_2O$) powder mixture at 12 GPa and 1673 K in 1 hour in the 3000-ton multi-anvil apparatus ORANGE-3000 at Geodynamics Research Center, Ehime University. Micro-focused XRD analyses (RAPIDII-V/DW) of the recovered sample showed that it consists mainly of stishovite (Stv), along with a minor $Al_2O_3$ phase that does not affect the EC measurements, as it is not interconnected within the sample. The $Al_2O_3$ content in Stv was determined to be 4.8 wt% by energy dispersive X-ray spectroscopy analyses. Observations with a field-emission-type scanning electron microscope (JSM-7000F) found that the hydrous Al-bearing $SiO_2$ sample exhibited homogeneous grain size distributions (~10 μm) (Figure 1). Water concentration was estimated by Fourier-transform infrared spectroscopy (FTIR) analyses (IRT-5200EUO) at Ehime University using the following equation (Paterson, 1982):

$$C_{\text{OH}} = \frac{X_\text{i}}{150\xi} \int \frac{K(v)}{(3780-v)} dv, \qquad (1)$$

where $C_{\text{OH}}$ is concentration of hydroxyl in ppm by weight, $\xi$ is orientation factor, $K(v)$ is absorption coefficient in $cm^{-1}$ for a given wavenumber $v$, and $X_\text{i}$ is a density factor. The FTIR spectra were collected with a $BaF_2$ window from three different locations of a sample under vacuum conditions by scanning 128 times per location in 100 × 100 μm² area (Figure 1). We employed the $X_\text{i} = 10^6 \times (18/2d)$ for Stv, in which $d$ is mineral density, giving $X_\text{i}$ = 2098 ppm $H_2O$ by weight (Bolfan-Casanova et al., 2000). The $H_2O$ content in the hydrous Al-bearing $SiO_2$ sample was determined to be 1750 ± 140 ppm by weight.

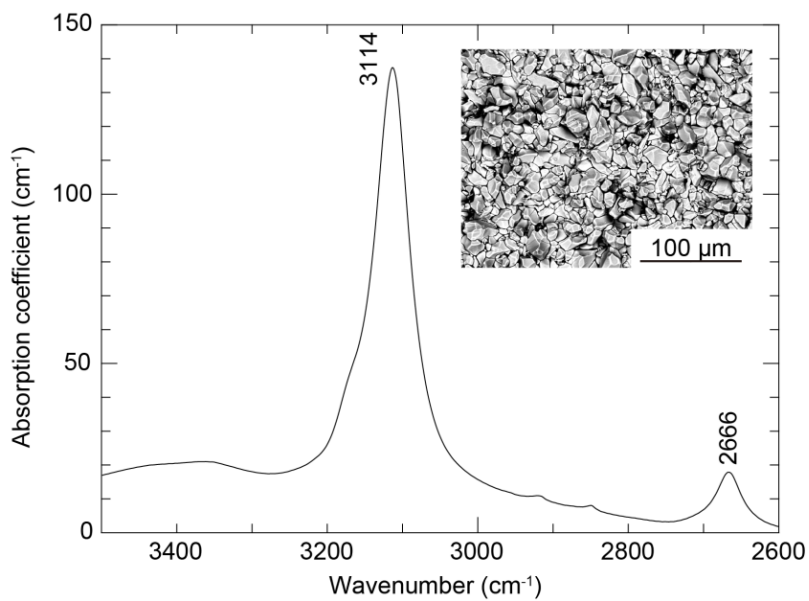

**Figure 1.** Unpolarized FTIR spectrum of the hydrous Al-bearing SiO$_2$ sample. Distinct peaks at 3114 and 2666 cm$^{-1}$ were observed, consistent with previously reported OH bands in dense SiO$_2$ (Bolfan-Casanova et al., 2000; Purevjav et al., 2024). The inset shows a scanning electron microscope image of the hydrous Al-bearing SiO$_2$ sample with an average grain size of ~10 μm.

2.2. Cell configuration and electrical conductivity measurements

EC measurements were carried out in a laser-heated DAC. Diamonds with 300 μm or 200 μm culet diameters were used for anvils. We drilled a hole with a size the same as that of the anvil culet at the center of a pre-indented rhenium gasket with a thickness of ~30 μm. A mixture of cubic boron nitride (cBN) + epoxy resin or TiO$_2$ (binder) was filled into the hole and subsequently compressed to ~50 μm in thickness. After compaction, a hole with 100 μm (or 50 μm) diameter was drilled in the cBN mixture using a pulsed laser for a sample chamber. The hydrous Al-bearing SiO$_2$ powder sample was made into a ~30 μm thick plate and loaded into the sample chamber, being sandwiched by 1) iridium foils (99.8% purity, Nilaco Corp.) (~3 μm thick and ~50 μm wide) that acted as both a laser absorber and an electrical resistance probe and 2) the ZrO$_2$ thermal insulation layers (10 μm thick) (Figure 2). Gold (~500 nm thick) was sputtered on the surface of the iridium foil such that the sample was not in direct contact with iridium. Gold and iridium, both platinum group metals, were used as electrode materials instead of commonly used platinum to prevent possible chemical reactions with hydrogen originating from the sample, leading to sample dehydration. This choice was supported by additional experiments, which suggested that platinum electrodes induce sample dehydration (Figure S1). The iridium foils were connected to the platinum leads placed on both sides of the gasket, which was further connected to copper lead wires, and a source meter (or an impedance analyzer). After loading, a whole DAC was placed in a vacuum oven at 393 K for 1 hour and the sample chamber was sealed while hot. Pressure was obtained by Raman spectroscopy measurements of a diamond anvil (Akahama & Kawamura, 2004). The effect of thermal pressure was added by employing the empirical relation of a 5% increase per 1000 K (Nomura et al., 2014).

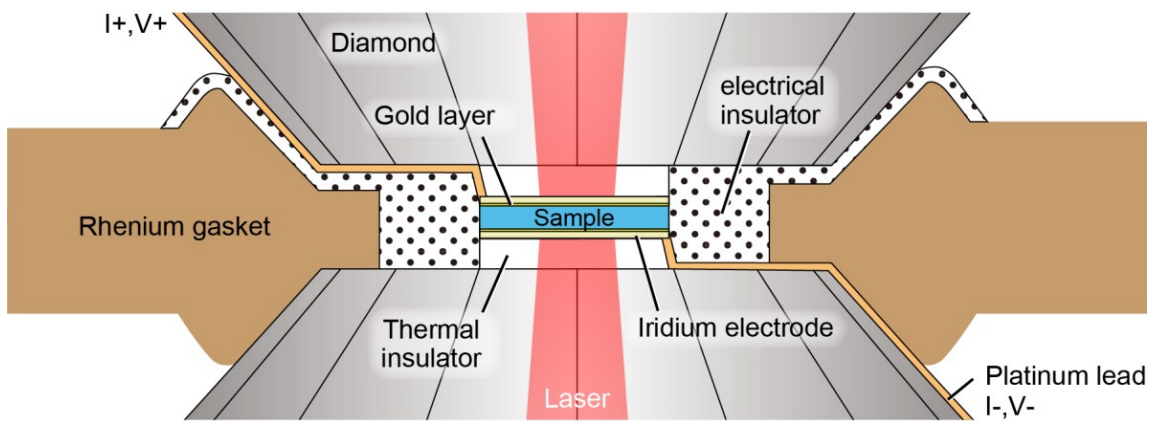

**Figure 2**. Schematic illustration of the cell assembly for electrical conductivity measurements. Platinum leads were connected to copper lead wires and a source meter (or an impedance analyzer).

The electrical resistance was measured under high *P-T* at the University of Tokyo. A double-sided laser heating system with a couple of 100 W single-mode Yb fiber lasers (YLR-100, IPG Photonics) was used to generate high temperatures in a DAC. We employed beam-shaping optics to produce a flat energy distribution. Temperature at the surface of the iridium electrode was determined by a spectro-radiometric method (Oka et al., 2019). The laser spot size ranged from 13 to 31 μm (Table S1). At laser output powers where the spectra were of insufficient quality for reliable temperature determination, the surface temperatures were estimated by interpolating the relationship between laser output power and temperature obtained from spectro-radiometric measurements (Dataset S1). The sample temperature between the two iridium electrodes was lower than the measured temperature at the electrode surface, and so the sample temperature was calculated by a finite element method using a software package COMSOL Multiphysics (COMSOL Inc.), as shown in Figure 3. For each experiment, temperature distributions were calculated based on the measured laser spot size and sample thickness. We employed a 3D model for the simulation and estimated a three-dimensional average temperature of a heated sample. We also confirmed that the results are consistent with those obtained using an axisymmetric 2D model (Figure S2). The difference between the temperature measured at the electrode surface and the simulated lowest sample temperature is only a few percent (Table S1) while the overall temperature uncertainty was considered to be ±10%.

In addition to these electrical resistance measurements, we performed independent X-ray diffraction (XRD) studies on the same hydrous Al-bearing $SiO_2$ sample at BL10XU, SPring-8 (Hirao et al., 2020) using a monochromatic X-ray beam of ~30 keV. The hydrous Al-bearing $SiO_2$ sample was mixed with fine-grained gold powder, serving as both a laser absorber and pressure marker, in a volume ratio of 10:1 using a mortar to ensure homogeneous mixing. The mixture of the sample and gold was then made into a pellet and loaded into a sample chamber, sandwiched between the layers of KCl serving as the pressure medium and the thermal insulator. We obtained the lattice volume and constants of the sample at 0–73 GPa and 300 K, each time after thermally annealing the sample with a laser. XRD data showed that the hydrous Al-bearing $SiO_2$ sample employed in this study has the orthorhombic $CaCl_2$-type (distorted stishovite) structure under high-pressure conditions of the present EC measurements (Figure 4).

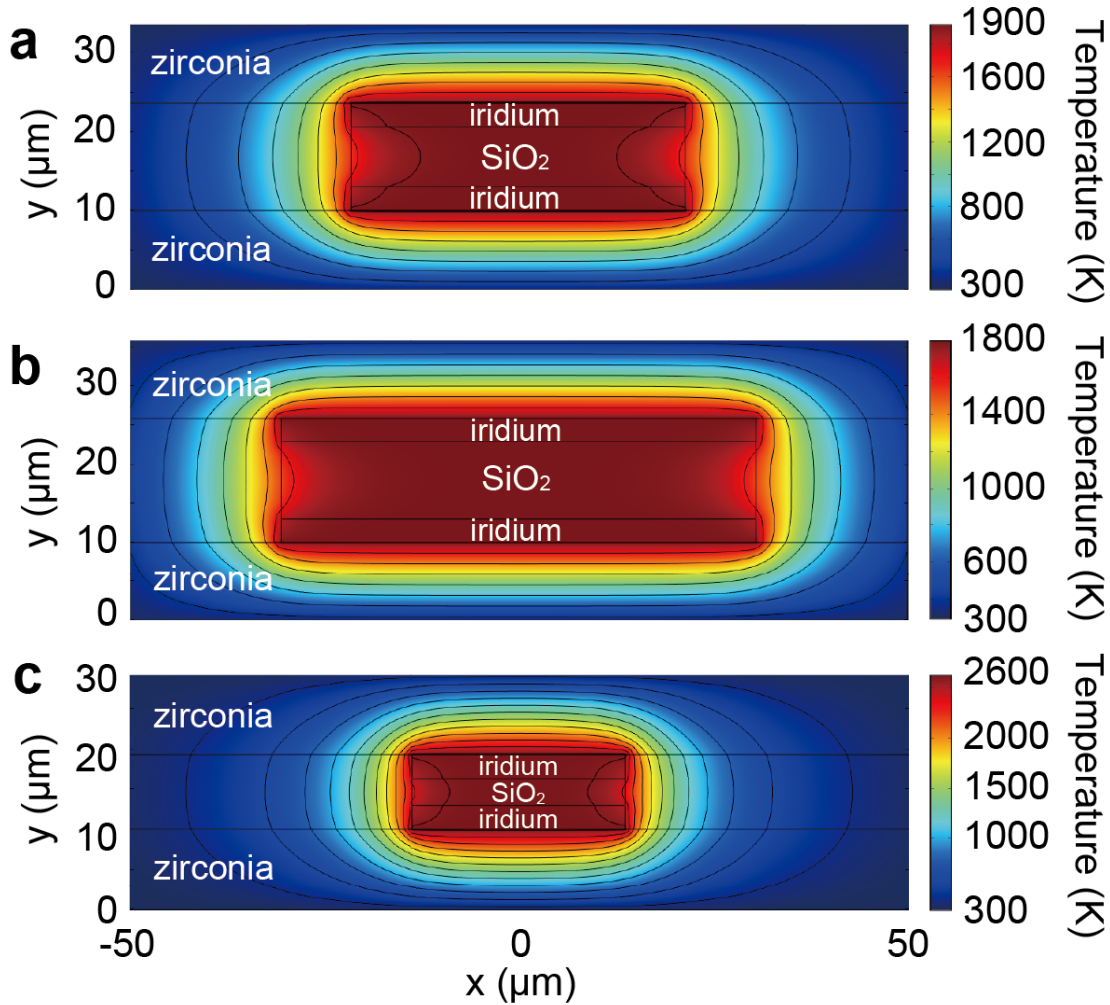

**Figure 3.** Simulated 2D cross-sectional temperature distributions of hydrous Al-bearing $SiO_2$ in (a) run #1 at 46 GPa heated to an average temperature of 1890 K, (b) run #2 at 57 GPa heated to an average temperature of 1800 K, and (c) run #3 at 82 GPa heated to an average temperature of 2610 K. Black curves indicate isotherms with 200 K intervals. A homogeneous circular area with a diameter of the laser spot size was considered on both sample surfaces, and the applied heat was conducted to the sample and surrounding 10 μm-thick zirconia layers. The back surface of the diamond anvils in contact with air and WC seats was set at room temperature. The thermal conductivity of $SiO_2$ was taken from high *P-T* computational study results (Aramberri et al., 2017), and that of zirconia was assumed to be a typical value for corundum under high *P-T* conditions and was set at 10 W/m/K (Hofmeister, 1999). Tetrahedral meshes with a resolution of 0.1 μm were applied.

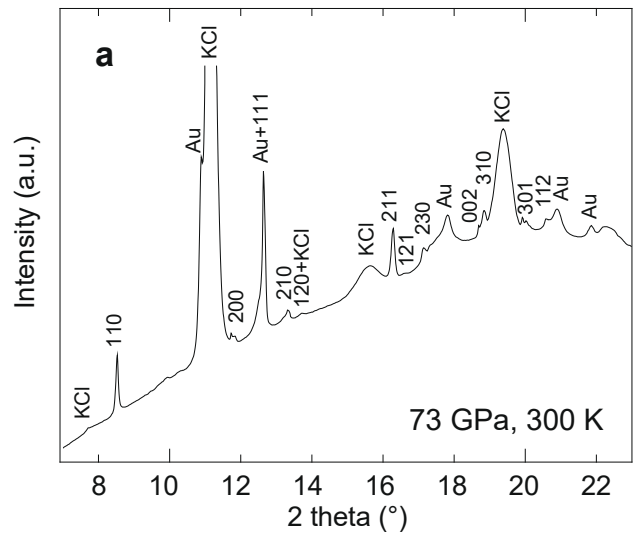
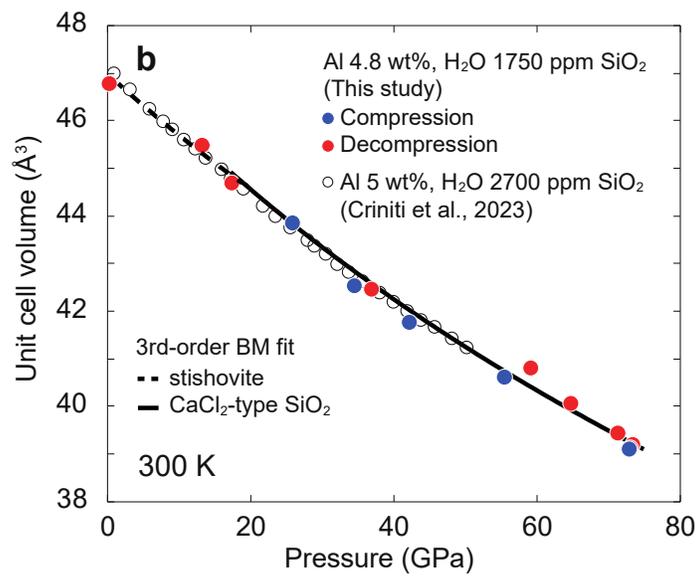
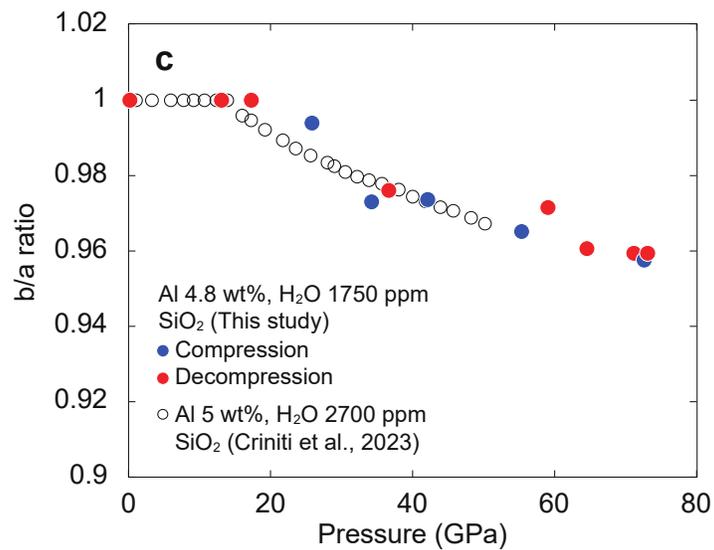

**Figure 4**. (a) Typical XRD spectrum collected at 73 GPa and 300 K. Numbers indicate Miller indices of the CaCl$_2$-type SiO$_2$ phase. (b) Pressure-volume curve at 300 K. Open circles indicate the previously reported compression data of hydrous Al-bearing SiO$_2$ (Criniti et al., 2023). Red and blue circles show data collected during compression and decompression, respectively. Broken and solid curves show 3$^{rd}$-order Birch-Murnaghan fitting results of tetragonal (stishovite) and orthorhombic (CaCl$_2$-type) SiO$_2$, respectively, which yields $K_0$ = 344(67) GPa, $K'$ = 4 (fixed), and $V_0$ = 46.9(3) Å$^3$ for stishovite, and $K_0$ = 245(14) GPa, $K'$ = 3.95(32), and $V_0$ = 48.0(18) Å$^3$ for the CaCl$_2$-type phase. (c) Change in the *b/a* axial ratio of our hydrous Al-bearing SiO$_2$ sample. The onset of the deviation of the *b/a* axial ratio from 1.0 above ~20 GPa indicates the transition from stishovite to the orthorhombic CaCl$_2$-type structure in our sample.

The electrical resistance of the hydrous Al-bearing SiO$_2$ sample measured in this study was primarily of direct-current (DC) resistance (Figure S3), in which heating duration was limited to about 1 second for a given *P-T* condition in order to avoid potential chemical reactions between the gold-sputtered iridium electrodes and the sample. The voltage of the sample was measured using a Keithley 2450 SourceMeter (Keithley Corp.) under a constant DC of 1 mA with 0.012% uncertainty in a pseudo-four-terminal assembly (Figure S4). The sample resistance was calculated from the applied current and the measured voltage via Ohm's law. To omit the effect of the Seebeck voltage possibly caused by a temperature heterogeneity in a sample, we obtained sample resistance by reversing the direction of the applied current every 1 ms (Ohta et al., 2023) (see Supplementary text S1 for details). In each heating cycle, the resistance data were collected ~1,000 times during heating for ~1 second and averaged to obtain the sample resistance. The relative uncertainty in sample resistance, $u_R$, was estimated as the standard deviation of the repeatedly collected sample resistance divided by its average value (Dataset S1).

In runs #2 and #3, impedance spectra were collected to determine the sample resistance without the effect of other source of resistance, such as the contact resistance, using a chemical impedance analyser HIOKI3532-80 and Four-Terminal Probe 9500. The alternative voltage was set to be 5 V, and the frequency range to be 10$^2$–10$^6$ Hz. The sample was continuously laser-heated while impedance spectra were collected. The typical duration for collecting a single impedance spectrum was around 30 seconds at each temperature. Although the heating time was longer than the DC measurements, no evidence of chemical reactions between the sample and electrodes was observed, as supported by the consistency between the results obtained from DC resistance and impedance measurements (see Results). The impedance of the system $Z(f) = R(f) + iX(f)$, where $R(f)$ and $X(f)$ are real and imaginary terms, respectively, are given in a Cole-Cole plot (Figure 5). We obtained the sample resistance by fitting a parallel circuit of resistance *R* and constant phase element to each spectrum using the commercial software Zview (Scribner Associates Inc.) and the following equation:

$$Z = \frac{R}{1+RC(2\pi i f)^p}, \quad (2)$$

where *C*, *f*, *i*, and *p* are capacitance, frequency, imaginary unit, and a fitting parameter ranging 0 < *p* ≤ 1 (*p* = 1 indicates an ideal capacitor), respectively.

The determination of EC ($\sigma$) from measured electrical resistance requires the sample thickness ($L$) and the cross-sectional surface area ($S$) as follows:

$$\sigma = \frac{L}{RS}. \qquad (3)$$

The thickness of the hydrous Al-bearing $SiO_2$ sample was measured on its cross section that was prepared with a focused Ga ion beam (FEI, Versa 3D DualBeam) after the sample recovery from a DAC (Figure S5). The sample thickness under high pressure was then estimated assuming isotropic sample volume change based on the equation of state (EoS) determined in this study for the present hydrous Al-bearing $SiO_2$ sample (Figure 4). The effect of volume expansion upon heating was ignored as it was shown to have a negligible effect (Figure S6). Regarding the surface area, we assumed that the electric current induced by an applied electric field flows only in a laser-heated portion. The $S$ value at high temperatures was therefore considered identical to the laser spot size (Table S1). The overall relative uncertainty in EC, $u_{EC}$, was calculated as follows:

$$u_{EC} = \sqrt{u_L^2 + u_S^2 + u_R^2} \qquad (4)$$

in which $u_L$ is the standard deviation of the observed sample thickness, $u_S$ is relative uncertainty in the surface area, and $u_R$ is fitting error of the electrical resistance. We found $u_{EC}$ to be typically ~20% (Dataset S1).

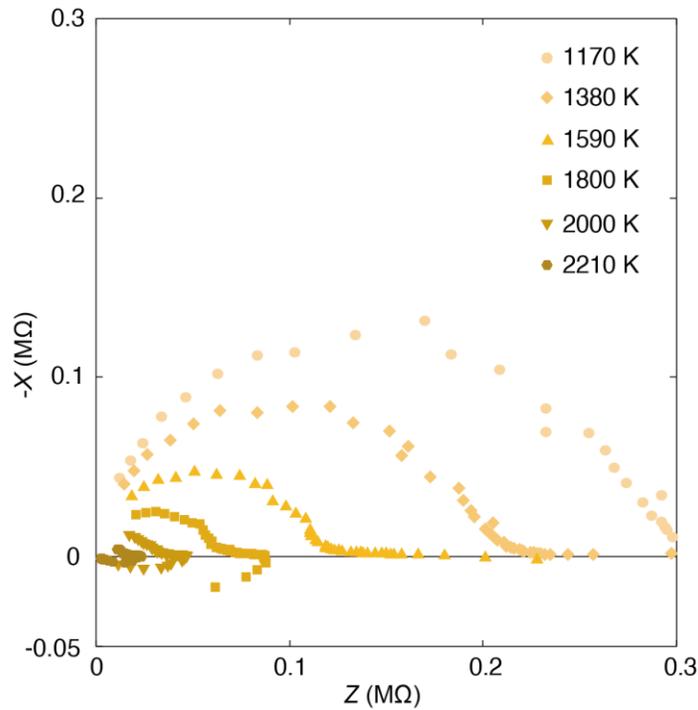

**Figure 5.** High *P-T* impedance spectra collected in the frequency range of $10^2$–$10^6$ Hz at 1170–2210 K in run #3. A low-frequency loop and a tail-like arc were observed in the collected impedance spectra at high temperatures, which might be due to 1) a high polarization resistance caused by a high voltage input (5 V) in our impedance spectroscopy (Keddam, 1984), or 2) electrode-sample reaction due to the long duration of laser-heating during impedance spectra acquisition (Barsoukov & Macdonald, 2018). We considered that such a low-frequency

arm was not derived from the sample; hence, we omitted it from the fitting process to estimate the sample resistance.

## 3. Results

In run #1, hydrous Al-bearing $SiO_2$ was compressed to 39 GPa at room temperature. Subsequently, we heated the sample to a certain temperature with a given laser power output, measured sample resistance during heating, and then quenched temperature in a few seconds (Dataset S1). Such a heating cycle was repeated with increasing laser power output up to 35 W in ~3 W intervals, which led to an increase in sample temperature to 1890 K at the maximum. Next, we repeated such short-time heating and resistance measurements by reducing the laser power output to 15 W in ~3 W intervals. The sample resistance could be obtained only down to $7 \times 10^6$ Ω at 960 K. The EC of hydrous Al-bearing $SiO_2$ ranged from ~$10^{-3}$ to $10^1$ S/m in this first run (Figure 6a). We found that the EC exhibited a relatively large positive temperature dependence below ~1200 K that decreased at high temperatures. We note that the obtained temperature-EC path was similar between heating and cooling cycles, which suggests the sample did not dehydrate upon heating.

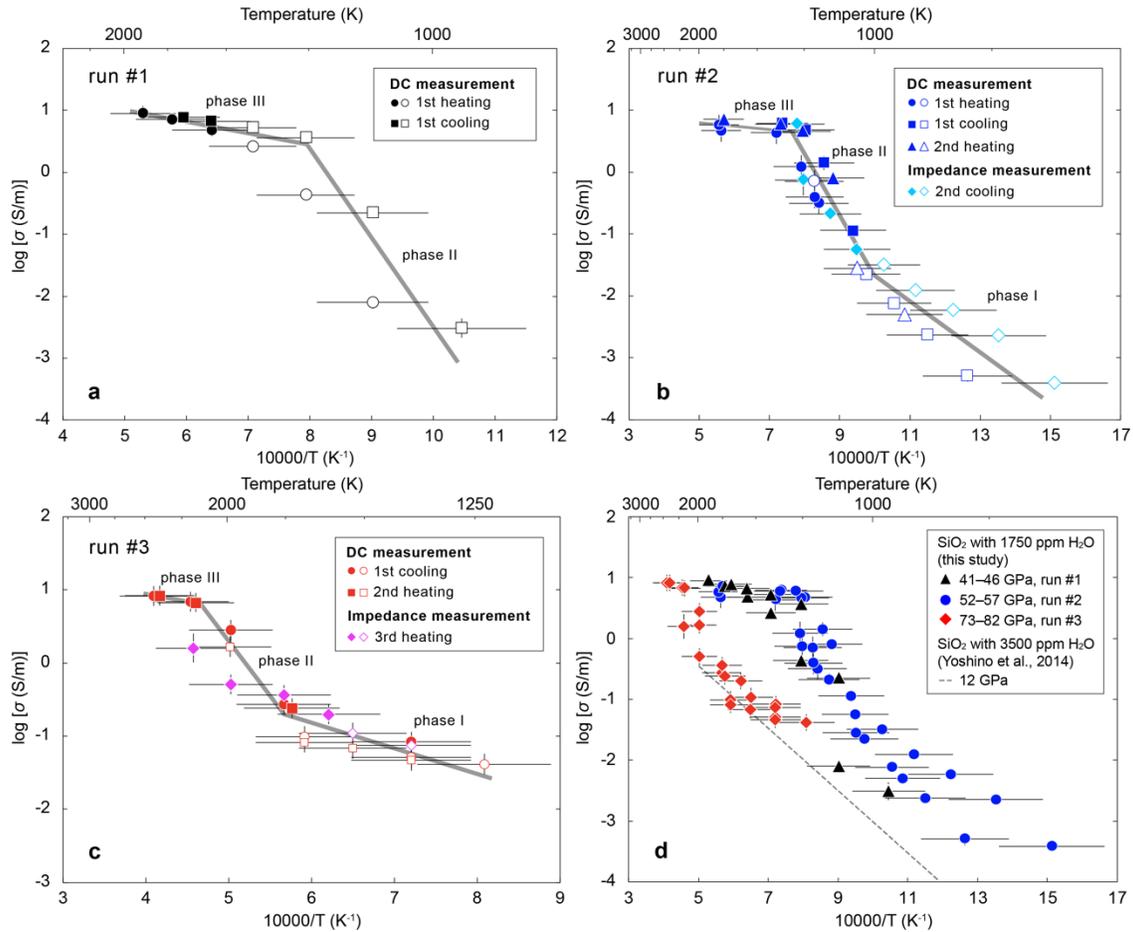

**Figure 6**. Electrical conductivity of Al-bearing $SiO_2$ with 1750 ppm $H_2O$ at (a) 41–46 GPa in run #1, (b) 52–57 GPa in run #2, and (c) 73–82 GPa in run #3. Open symbols indicate data where interpolated temperature values were used. (d) Plots of all data. Triangles, run #1; circles, run #2; diamonds, run #3. Gray broken line indicates earlier data obtained at 12 GPa in a multi-anvil press for $SiO_2$ stishovite with 3500 ppm $H_2O$ (Yoshino et al., 2014). The conductivity difference between the previous measurements and this study may be attributed to the difference in the crystal structure of $SiO_2$.

The second experiment was carried out at ~54 GPa (Dataset S1). Sample resistance was measured during both heating and cooling, similar to run #1, with the additional use of impedance spectroscopy. The data collected in a wide temperature range, including relatively low temperatures, indicate that the EC exhibited moderate temperature dependence below ~1000 K (phase I), strong dependence between ~1000 K and ~1250 K (phase II), and weak dependence above ~1250 K (phase III) (Figure 6b). The EC reached ~10 S/m at ~1250 K and remained nearly constant up to 1800 K, consistent with observations made in run #1. The EC determined from the impedance spectra collected during the second cooling was found to be consistent with that obtained from the DC measurements, indicating that both the grain boundary resistance and the electrode interface resistance are negligible compared to the sample resistance. The temperature–EC relation obtained in run #2 at ~54 GPa closely overlaps with that of run #1 at ~44 GPa above ~1200 K, likely because the room-temperature EC is higher while the temperature dependence is smaller at higher pressure.

In run #3 conducted at ~78 GPa, the temperature–EC relation was again remarkably different among three temperature ranges; <~1600 K (moderate increase, phase I), ~1600 to ~2200 K (rapid increase, phase II) and >~2200 K (least increase, phase III) (Figure 6c). The EC data obtained by the impedance measurements agree with those from the DC resistance data, similar to run #2. The temperature ranges of the phases II and III were much higher in this experiment than those observed in run #2 at ~54 GPa (Figure 6d). On the other hand, the EC changed little in phase III once it reached ~10 S/m in all of these three separate runs performed at different pressure ranges. The least pressure and temperature dependence of the EC has also been observed for dense ionic $H_2O$ fluid with ~10–40 S/m at 11–45 GPa and 1030–2750 K (Oka et al., 2024). The high EC of ~10 S/m was not found in previous measurements on hydrous $SiO_2$ to 1900 K at 12 GPa (Yoshino et al., 2014).

**4. Discussion**

4.1. Transition to superionic state in hydrous Al-bearing $SiO_2$

The temperature dependence of EC, $\sigma$, at a given pressure, is written as:
$$\sigma = \sigma_0 \exp(-\Delta H/kT), \tag{5}$$
where $\sigma_0$ is a pre-exponential factor, $\Delta H$ is activation enthalpy, and k is the Boltzmann constant. By fitting this equation respectively to the present EC data collected in each run at the low-temperature range (phase I), we obtained $\Delta H$ = 0.67 eV at ~54 GPa and 0.46 eV at ~81 GPa (Table S2), suggesting that the dominant conduction mechanism is proton conduction. The extrapolation of the present $\Delta H$ value to 12 GPa by assuming linear pressure dependence yielded 1.00 eV, which is consistent with the reported Δ*H* value for hydrous stishovite with 3500 ppm $H_2O$ (Yoshino et al., 2014).

It has been predicted that the proton diffusion in hydrous Al-bearing $SiO_2$ stishovite (or $CaCl_2$-type $SiO_2$ phase) is gradually enhanced with increasing temperature and eventually becomes superionic (Umemoto et al., 2016). In superionic state, protons become extremely diffusive like a fluid, and thus their mobility shows weak temperature dependence. Indeed, a recent theoretical study demonstrated that a transition to the superionic state in $H_2O$ ice reduces Δ*H* from 0.98 to 0.24 eV, accompanying an increase in EC by two orders of magnitude (Sun et al., 2020). Phase III in this work exhibited high EC, about two orders of magnitude greater than phase I, and relatively small Δ*H* values ranging from 0.08–0.35 eV (Table S2) that may correspond to a superionic transition. Moreover, the occurrence of phase II showing the abrupt increase in the EC and its temperature dependence agrees with the predicted behaviour of a superionic transition in hydrous Al-bearing $SiO_2$ (Umemoto et al., 2016). Our results suggest the sample becomes fully superionic (phase III) above 1110 K at 42 GPa, 1200 K at 55 GPa, and 2190 K at 79 GPa, indicating an increasing transition temperature to the fully superionic state with increasing pressure (Figure 7).

Note that our sample is highly unlikely to have dehydrated during the measurement; the $H_2O$ capacity of Al-free $SiO_2$ is limited at high temperatures (Takaichi et al., 2024), whereas Al-bearing $SiO_2$ demonstrates a significantly higher capacity, exceeding 1 wt% at temperatures above 4000 K (Tsutsumi et al., 2024). The $H_2O$ capacity of Al-bearing $SiO_2$ in the investigated *P-T* range was reported to be 0.5–1.5 wt% (Ishii et al., 2022; Tsutsumi et al., 2024), which is exceedingly higher than the $H_2O$ concentration in our sample of 0.175 wt%. We reemphasize that the consistency of the collected EC during the heating and cooling cycles also supports the absence of sample dehydration in this study.

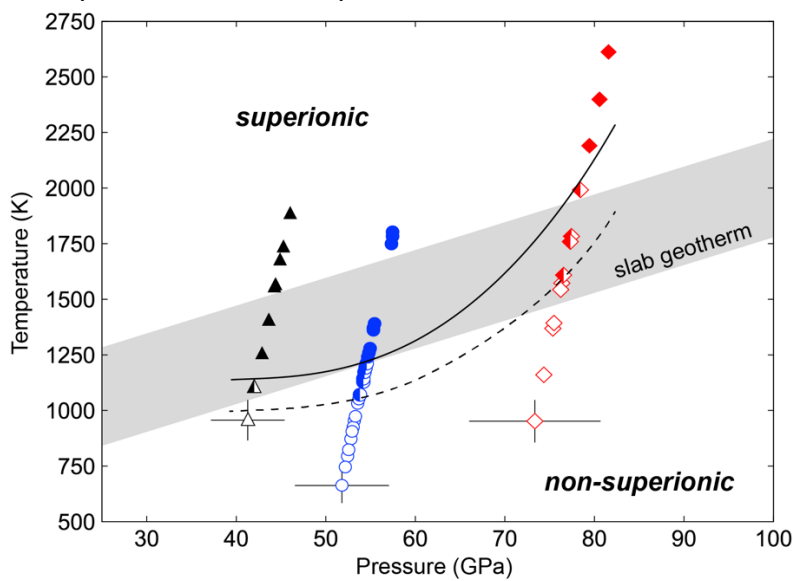

**Figure 7**. Pressure and temperature conditions of the superionic transition in hydrous Al-bearing $SiO_2$. Black triangle, run #1; blue circle, run #2; red diamond, run #3. Broken and solid curves indicate transitions from phase I to II (non-superionic to partially superionic) and from phase II and III (fully superionic), respectively. Open, half-filled, and closed symbols indicate data corresponding to observations of phases I, II, and III, respectively. Error bars are given only to data at the lowest temperature in each run. The gray band indicates the slab geotherm (Ohtani, 2020).

The proton diffusion coefficient (*D*) can be estimated from our measured EC values using the following equation:

$$D = \frac{\sigma_p RT}{nz^2 F^2}, \tag{6}$$

where $\sigma_p$ is proton conductivity, R is the gas constant, *T* is absolute temperature, *n* is the number density of protons in mol/m$^3$, *z* is valency (proton: z=1), and F is Faraday's constant. Figure 8 shows the estimated logarithm of the proton diffusion coefficient in hydrous Al-bearing $SiO_2$ as a function of reciprocal temperature. As seen in the EC data, the temperature dependence of proton diffusion coefficient shows three distinct trends: an initial moderate increase, followed by a rapid increase, and finally a minimal increase as the reciprocal temperature decreases. For instance, at 73–82 GPa (run #3), the logarithm of the proton diffusion coefficient increased linearly with decreasing reciprocal temperature. The diffusion coefficient increased from $10^{-6.2}$ cm$^2$/s at 1160 K to $10^{-5.8}$ cm$^2$/s at 1570 K. This was followed by a sharp increase to $10^{-3.8}$ cm$^2$/s at 2200 K, corresponding to the transition to a superionic state, and subsequently a slight increase to $10^{-3.6}$ cm$^2$/s as the temperature was increased further to 2610 K. The proton diffusion coefficients in superionic hydrous Al-bearing $SiO_2$ exhibited such significantly high values (~$10^{-4}$ cm$^2$/s) that are comparable to those of superionic $H_2O$ ice XVIII (Sun et al., 2020). This confirms that the high proton diffusion and EC of hydrous Al-bearing $SiO_2$ at elevated temperatures characterize the fully superionic state at phase III. The difference in the temperature dependencies of the proton diffusion coefficient between superionic $H_2O$ ice XVIII and hydrous Al-bearing $SiO_2$, particularly the stronger temperature dependence in the former, may be attributed to a difference in crystal structure or pressure effects. As liquids generally exhibit a positive activation volume, the temperature dependence of the EC and the diffusion coefficient (Δ*H*) typically increases with pressure (Zhang et al., 2021). Since protons in superionic $H_2O$ ice and hydrous Al-bearing $SiO_2$ are expected to exhibit liquid-like behavior, it is likely that their Δ*H* also shows a positive pressure dependence. This could account for the observed high Δ*H* in superionic $H_2O$ ice XVIII at 200 GPa compared to superionic hydrous Al-bearing $SiO_2$ at 41–82 GPa (Figure 8).

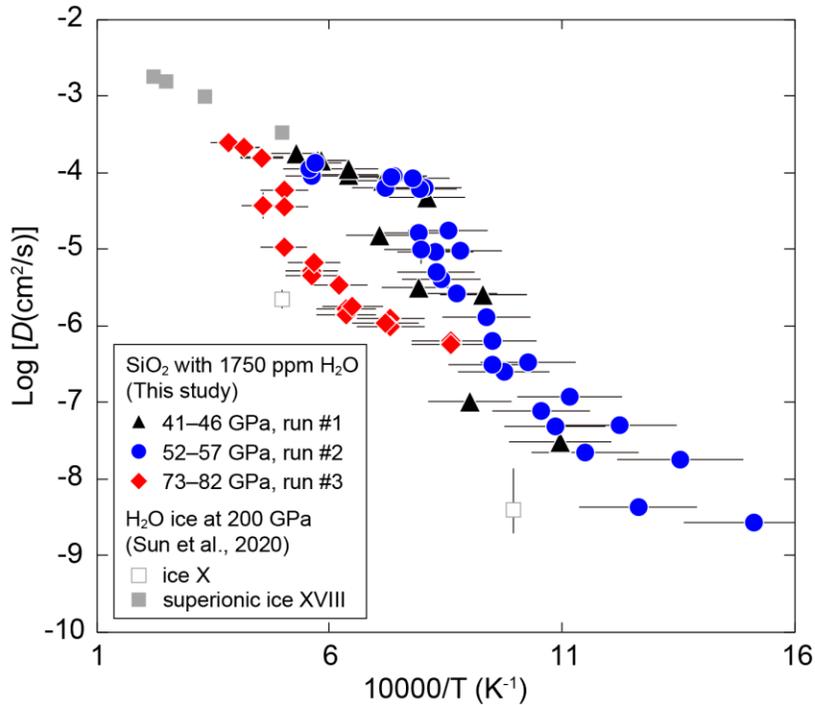

**Figure 8**. Logarithm of proton diffusion coefficient of hydrous Al-bearing SiO$_2$ as a function of reciprocal temperature. Black triangles, run #1; blue circles, run #2; red diamonds, run #3. Open and closed gray squares are proton diffusion coefficients of H$_2$O ice X and superionic ice XVIII, respectively, at 200 GPa and high temperatures estimated from *ab initio* molecular dynamics simulations (Sun et al., 2020).

4.2. Electrical conductivity of subducted MORB crusts

The high *P-T* conditions of superionic hydrous Al-bearing SiO$_2$ overlap with those of cold subducting plates (slabs) in the shallow to middle lower mantle (Figure 7). The temperatures of the subducting plates are not high enough for the fully superionic state above ~75 GPa, corresponding to 1800 km depth, although the partially superionic conduction continues to ~85 GPa. SiO$_2$ stishovite or the CaCl$_2$-type phase in the slab crusts includes up to ~3 wt% H$_2$O at 700 to 1800 km depth (Ishii et al., 2022; Lin et al., 2022; Tsutsumi et al., 2024). While our measurements show the EC of superionic Al-bearing SiO$_2$ containing 1750 ppm (0.175 wt%) H$_2$O to be ~10 S/m, it is likely to be as high as ~160 S/m when the SiO$_2$ phase includes 3 wt% H$_2$O (Figure 9a), since the EC of SiO$_2$ via proton conduction is known to be proportional to its H$_2$O content, which is predicted by Nernst-Einstein equation (Yoshino, 2010) and has been confirmed experimentally up to 0.26 wt% H$_2$O (Yoshino et al., 2014).

The EC of superionic Al-bearing SiO$_2$ containing 1750 ppm H$_2$O was found to be more than an order of magnitude higher than that of anhydrous MORB comprising majorite with 10–15 vol% SiO$_2$ at *P-T* conditions equivalent to those of the topmost lower mantle (Yoshino et al., 2008) (Figure 9a). At higher pressures, the EC of anhydrous MORB containing $SiO_2$ stishovite has been measured at a single pressure of 51 GPa between 1430 K and 2100 K (Ohta et al., 2010). Fitting equation 5 to their data yields $\sigma_0$ = 66.0 S/m, and $\Delta H$ (= $\Delta E + P\Delta V$) = 0.376 eV at this pressure. On the other hand, activation volume $\Delta V$, which gives the pressure dependence of $\Delta H$, was not obtained from their data (Ohta et al., 2010). Alternatively, we assumed $\Delta V$ = -0.55 cm$^3$/mol for MORB to be the same as that for (Fe,Al)-bearing bridgmanite (Sinmyo et al., 2014), a primary mineral in the subducted MORB material, which yielded $\Delta E$ = 0.379 eV for the dry MORB. By applying $\sigma_0$ = 66.0 S/m, $\Delta H$ at a given pressure calculated with such $\Delta E$ and $\Delta V$ values, and $T$ along a typical cold slab geotherm (Ohtani, 2020) in Equation 5, we estimated the EC of the subducted dry MORB crust to be 0.5–5.0 S/m at 660–1800 km depth (Figure 9a). It matches the averaged one-dimensional depth profile of the EC estimated from geomagnetic observations (Velímský & Knopp, 2021) at ~800–1200 km depth but is lower than observed in the shallower lower mantle (660–800 km depth). The present measurements demonstrate ~10 S/m for superionic Al-bearing $SiO_2$ containing 0.175 wt%, with extrapolation to 3 wt% $H_2O$ yielding a conductivity of ~160 S/m. Both values are much higher than the EC of anhydrous MORB, suggesting that subducted hydrous MORB materials exhibit higher EC (Figure S7).

The hydrous Al-bearing $SiO_2$ phase constitutes ~25% of the MORB assemblage by volume in the lower mantle (Ono et al., 2001). It possibly forms an interconnected film in subducted MORB crusts and governs its bulk EC. While dense dry $SiO_2$ is a hard crystal (Hunt et al., 2019), crystals may be significantly weakened by the presence of structural $H_2O$. For example, the hardness of dry $SiO_2$ quartz at 1300 K drops from ~2 GPa to ~0.2 GPa by adding 1300 ppm of $H_2O$ (Griggs, 1967). While there is no available report on the deformation of hydrous $SiO_2$ stishovite/$CaCl_2$-type phase, it has been repeatedly documented that $H_2O$ significantly weakens mantle minerals (Karato & Jung, 2003; Mei & Kohlstedt, 2000). Therefore, when it hosts a few wt% $H_2O$ (Ishii et al., 2022; Lin et al., 2022; Tsutsumi et al., 2024), it is feasible for $H_2O$-bearing $SiO_2$ to be substantially weaker than other coexisting phases in MORB, such as the most abundant mineral of bridgmanite primarily accommodating the strain during subduction (Girard et al., 2016) and forming an interconnected film. Accordingly, we calculated the EC of hydrous MORB including 25 vol% superionic Al-bearing $SiO_2$ phase with 3 wt% $H_2O$ based on Hashin-Shtrikman's perfectly connected model (Hashin & Shtrikman, 1963) and found $10^{1.4-1.6}$ S/m at 660–1800 km depth, which is one to two orders of magnitude higher than that of the anhydrous MORB (Figure 9a). We further evaluated the effect of hydration on MORB in the mantle transition zone and the uppermost lower mantle, where MORB consists of majorite garnet and 10–15% dense $SiO_2$ (Hirose et al., 1999). Our results show that MORB crusts including superionic Al-bearing $SiO_2$ with 3 wt% $H_2O$ show approximately an order of magnitude higher EC than the dry MORB case (Yoshino et al., 2008). Even when $SiO_2$ does not form an interconnected network, the conductivity still increased by approximately 30% (Figure S8). These findings highlight the important role of hydrous $SiO_2$ in enhancing the EC of subducted MORB crust.

Seismological observations and laboratory measurements of the sound velocity of MORB minerals suggest that MORB crusts are piled at the top of the lower mantle (Feng et al., 2021; Gréaux et al., 2019; Niu, 2014; Schmandt et al., 2014) (Figure 9b). This is consistent with the previous findings that subducted MORB crust becomes buoyant near the 660 km discontinuity (Irifune & Ringwood, 1993) and remains buoyant up to approximately 720 km depth (Hirose et al., 1999). Mantle convection simulations have also indicated that the MORB crust separates from the slab during subduction (Christensen & Hofmann, 1994; Nakagawa & Tackley, 2005), enabling the accumulation of MORB material at these depths. The present finding that the EC of the hydrous MORB assemblage containing superionic hydrous Al-bearing $SiO_2$ exceeds the high EC anomaly found underneath NE China offers an alternative to silicate melt or dry MORB, whose EC is lower than geomagnetic observations. While the transport of $H_2O$ by subducting slabs into the lower mantle has long been a matter of debate (Ohtani, 2020; Walter, 2021), the present study suggests that the high EC anomalies observed atop the lower mantle can be reconciled with piled hydrous MORB materials owing to the highly conductive superionic $SiO_2$ phase; the EC of MORB that includes Al-bearing $SiO_2$ containing ~0.2 wt% $H_2O$ matches the observed anomaly (Figure 9a).

In addition, seismological observations have also revealed notable mid-lower mantle seismic scattering beneath the Japan Sea at depths of 930–1120 km (Li & Yuen, 2014; Niu, 2014) (Figure 9b), which has been believed to be caused by the ferroelastic-type phase transition between stishovite and the $CaCl_2$-type phase in Al-bearing $SiO_2$ (Nomura et al., 2010) included in subducted MORB materials as a part of the ancient Izanagi plate (Li & Yuen, 2014). The EC locally observed in this region (Kelbert et al., 2009; Kuvshinov, 2012) exhibits a high anomaly, specifically at 900–1200 km depth, compared to that of the surrounding mantle and is consistent with that of the hydrous MORB with the superionic Al-bearing $SiO_2$ phase that contains 0.2 wt% $H_2O$ (Figure 9a; Figure S7). While both the Pacific plate and the ancient Izanagi plate represent cold plates and should have thus carried relatively large amounts of $H_2O$ in their crustal part during their subduction, such comparison between the EC of hydrous MORB and the geomagnetic observations suggests 0.2 wt% $H_2O$ in hydrous $SiO_2$, which is lower by one order of magnitude than its storage capacity (Lin et al., 2022). The present work demonstrates the capability of EC measurements at high *P-T* to unveil the distributions of $H_2O$ in the Earth's deep mantle beyond the mantle transition zone (Huang et al., 2005; Yoshino et al., 2008).

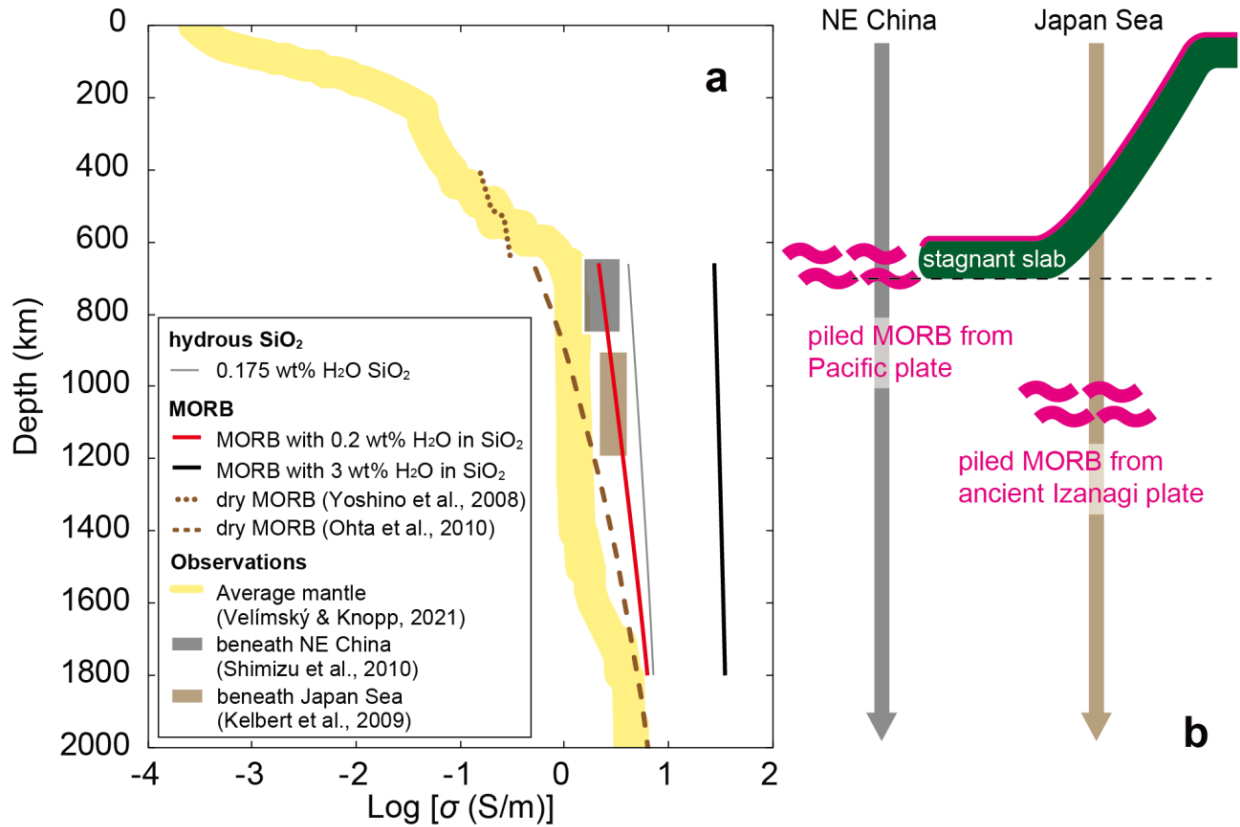

**Figure 9**. (a) Comparison of the electrical conductivities of dry and hydrous MORBs with global and local observations. The present measurements provide the electrical conductivity of superionic Al-bearing $SiO_2$ with 0.175 wt% $H_2O$ along the slab geotherm (Ohtani, 2020) (gray thin line). Red and black bold lines show the calculated electrical conductivity of hydrous MORB that includes the superionic Al-bearing $SiO_2$ phase with 0.2 wt% and 3 wt% $H_2O$, respectively. Brown dotted line represents the electrical conductivity of dry majoritic MORB (Yoshino et al., 2008). Brown dashed line gives the electrical conductivity estimated based on post-majorite MORB data (Ohta et al., 2010). Yellow band represents the globally-averaged, one-dimensional mantle electrical conductivity model based on geomagnetic observations (Velímský & Knopp, 2021). The gray and brown rectangles show the local observations of high electrical conductivity anomalies underneath NE China at 650–850 km depth (Kelbert et al., 2009; Kuvshinov, 2012; Shimizu et al., 2010) and Japan Sea at 900–1200 km depth (Kelbert et al., 2009; Kuvshinov, 2012), respectively. (b) Schematic illustrations of the slab beneath NE China and the Japan Sea. Broken line indicates the 660 km depth. Seismological observations reported the piled subducted MORB crusts in the shallow (Feng et al., 2021) to middle-lower mantle (Li & Yuen, 2014).

## 5. Conclusions

We determined the EC of CaCl$_2$-type Al-bearing SiO$_2$ containing 1750 ppm H$_2$O and 4.8 wt% Al$_2$O$_3$ at 41–82 GPa and 660–2610 K in a laser-heated DAC using a recently-developed technique to measure the EC of materials that are transparent to a near-infrared laser beam conventionally used for heating a sample in the DAC (Okuda et al., 2022). We identified three distinct EC regimes characterized by moderate (phase I), strong (phase II), and weak (phase III) temperature dependence, which we interpret as corresponding to proton conduction, the onset of superionic transition, and the fully superionic state, respectively. The transition temperature to the fully superionic state (phase III) increases with pressure, occurring above 1110 K at 42 GPa, 1200 K at 55 GPa, and 2190 K at 79 GPa. The *P-T* range of the superionicity of hydrous Al-bearing SiO$_2$ overlaps with that of subducting slabs in the Earth's lower mantle up to ~1800 km depth. The measured EC of superionic hydrous Al-bearing SiO$_2$ is approximately 10 S/m, several to ten times higher than that of the shallow to middle lower mantle, and could be even higher if more H$_2$O is included.

Al-bearing SiO$_2$ is one of major phases in subducted crustal materials, constituting ~25 vol% in subducted MORB crusts in the lower mantle. The hydrous Al-bearing SiO$_2$ phase is possibly substantially weaker than other coexisting phases in MORB, forming an interconnected film during subduction and predominating the EC of the bulk MORB materials. If this is the case, the EC of MORBs containing the superionic hydrous Al-bearing SiO$_2$ phase is estimated to be ~$10^{0.5}$ to ~$10^{1.5}$ S/m in the lower mantle to ~1800 km depth when 0.2 to 3 wt% H$_2$O is included in the SiO$_2$ phase (0.05 to 0.75 wt% H$_2$O in the bulk MORB). Such high EC of MORB may explain the high EC anomalies observed at subduction zones underneath NE China.

**Acknowledgments**

We thank K. Baba, H. Shimizu and S. Azuma for valuable discussion. Comments from two anonymous reviewers and the associate editor helped improve the manuscript. The hydrous SiO$_2$ sample was synthesized based on the PRIUS program at the GRC, Ehime University. This work was supported by JSPS KAKENHI (grant no. 21H04968). XRD measurements were carried out at the beamline BL10XU, SPring-8 under the approval with JASRI (proposal no. 2021B0181).

**Open Research**

Datasets for this research Dataset S1 are available at (Inada et al., 2025).

Gleason, A. E., Rittman, D. R., Bolme, C. A., Galtier, E., Lee, H. J., Granados, E., et al. (2022). Dynamic compression of water to conditions in ice giant interiors. *Scientific Reports*, *12*(1), 715. https://doi.org/10.1038/s41598-021-04687-6

Gréaux, S., Irifune, T., Higo, Y., Tange, Y., Arimoto, T., Liu, Z., & Yamada, A. (2019). Sound velocity of $CaSiO_3$ perovskite suggests the presence of basaltic crust in the Earth's lower mantle. *Nature*, *565*(7738), 218–221. https://doi.org/10.1038/s41586-018-0816-5

Griggs, D. (1967). Hydrolytic weakening of quartz and other silicates. *Geophysical Journal of the Royal Astronomical Society*, *14*(1–4), 19–31. https://doi.org/10.1111/j.1365-246X.1967.tb06218.x

Guo, X., & Yoshino, T. (2013). Electrical conductivity of dense hydrous magnesium silicates with implication for conductivity in the stagnant slab. *Earth and Planetary Science Letters*, *369–370*, 239–247. https://doi.org/10.1016/j.epsl.2013.03.026

Hashin, Z., & Shtrikman, S. (1963). A variational approach to the theory of the elastic behaviour of multiphase materials. *Journal of the Mechanics and Physics of Solids*, *11*(2), 127–140. https://doi.org/10.1016/0022-5096(63)90060-7

Hirao, N., Kawaguchi, S. I., Hirose, K., Shimizu, K., Ohtani, E., & Ohishi, Y. (2020). New developments in high-pressure X-ray diffraction beamline for diamond anvil cell at SPring-8. *Matter and Radiation at Extremes*, *5*(1), 018403. https://doi.org/10.1063/1.5126038

Hirose, K., Fei, Y., Ma, Y., & Mao, H.-K. (1999). The fate of subducted basaltic crust in the Earth's lower mantle. *Nature*, *397*(6714), 53–56. https://doi.org/10.1038/16225

Hirose, K., Takafuji, N., Sata, N., & Ohishi, Y. (2005). Phase transition and density of subducted MORB crust in the lower mantle. *Earth and Planetary Science Letters*, *237*(1–2), 239–251. https://doi.org/10.1016/j.epsl.2005.06.035

Hofmeister, A. M. (1999). Mantle values of thermal conductivity and the geotherm from phonon lifetimes. *Science*, *283*(5408), 1699–1706. https://doi.org/10.1126/science.283.5408.1699

Hou, M., He, Y., Jang, B. G., Sun, S., Zhuang, Y., Deng, L., et al. (2021). Superionic iron oxide–hydroxide in Earth's deep mantle. *Nature Geoscience*, *14*(3), 174–178. https://doi.org/10.1038/s41561-021-00696-2

Huang, X., Xu, Y., & Karato, S. (2005). Water content in the transition zone from electrical conductivity of wadsleyite and ringwoodite. *Nature*, *434*(7034), 746–749. https://doi.org/10.1038/nature03426
21

Komabayashi, T., Maruyama, S., & Rino, S. (2009). A speculation on the structure of the D″ layer: the growth of anti-crust at the core–mantle boundary through the subduction history of the Earth. *Gondwana Research*, *15*(3–4), 342–353. https://doi.org/10.1016/j.gr.2008.11.006

Koyama, T., Shimizu, H., Utada, H., Ichiki, M., Ohtani, E., & Hae, R. (2006). Water content in the mantle transition zone beneath the north Pacific derived from the electrical conductivity anomaly. In S. D. Jacobsen & S. Van Der Lee (Eds.), *Geophys. Monogr. Ser.* (pp. 171–179). Washington, D. C.: American Geophysical Union. Retrieved from https://onlinelibrary.wiley.com/doi/10.1029/168GM13

Kuvshinov, A. V. (2012). Deep electromagnetic studies from land, sea, and space: progress status in the past 10 years. *Surveys in Geophysics*, *33*(1), 169–209. https://doi.org/10.1007/s10712-011-9118-2

Li, Juan, & Yuen, D. A. (2014). Mid-mantle heterogeneities associated with izanagi plate: implications for regional mantle viscosity. *Earth and Planetary Science Letters*, *385*, 137–144. https://doi.org/10.1016/j.epsl.2013.10.042

Li, Junwei, Lin, Y., Meier, T., Liu, Z., Yang, W., Mao, H., et al. (2023). Silica-water superstructure and one-dimensional superionic conduit in Earth's mantle. *Science Advances*, *9*(35), eadh3784. https://doi.org/10.1126/sciadv.adh3784

Lin, Y., Hu, Q., Walter, M. J., Yang, J., Meng, Y., Feng, X., et al. (2022). Hydrous $SiO_2$ in subducted oceanic crust and $H_2O$ transport to the core-mantle boundary. *Earth and Planetary Science Letters*, *594*, 117708. https://doi.org/10.1016/j.epsl.2022.117708

Manthilake, G., Schiavi, F., Zhao, C., Mookherjee, M., Bouhifd, M. A., & Jouffret, L. (2020). The electrical conductivity of liebermannite: implications for water transport into the Earth's lower mantle. *Journal of Geophysical Research: Solid Earth*, *125*(8), e2020JB020094. https://doi.org/10.1029/2020JB020094

Mei, S., & Kohlstedt, D. L. (2000). Influence of water on plastic deformation of olivine aggregates: 2. Dislocation creep regime. *Journal of Geophysical Research: Solid Earth*, *105*(B9), 21471–21481. https://doi.org/10.1029/2000JB900180

Millot, M., Hamel, S., Rygg, J. R., Celliers, P. M., Collins, G. W., Coppari, F., et al. (2018). Experimental evidence for superionic water ice using shock compression. *Nature Physics*, *14*(3), 297–302. https://doi.org/10.1038/s41567-017-0017-4
23

**Supporting Information**

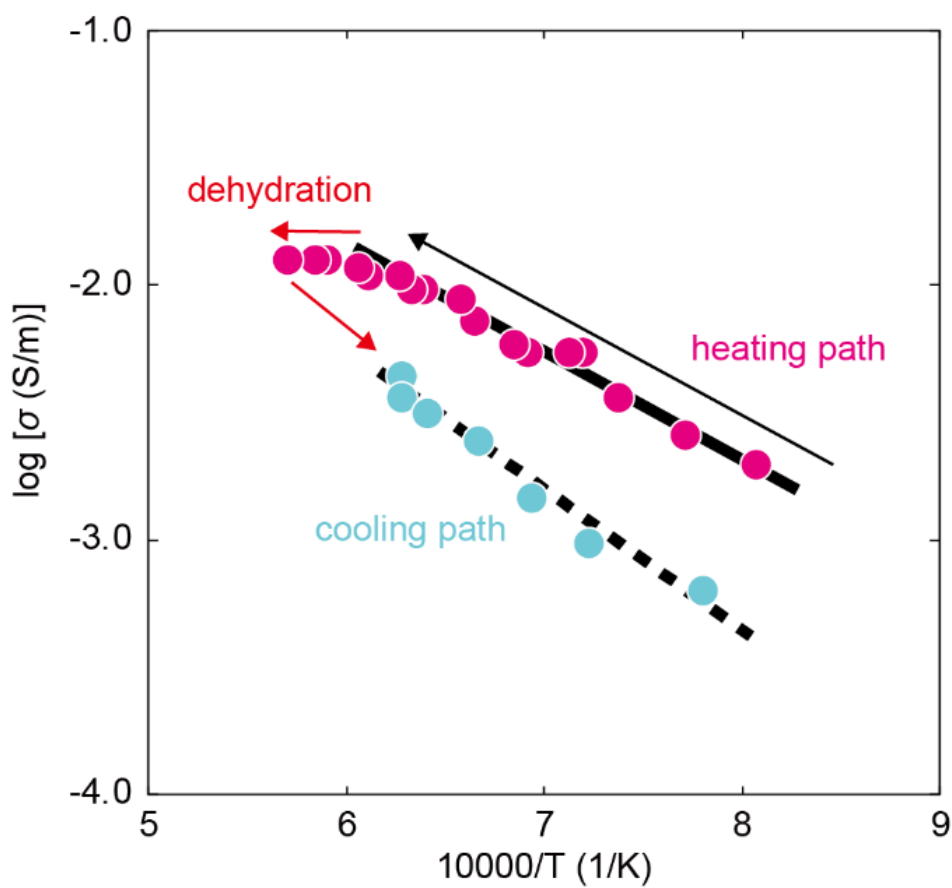

**Figure S1.** Electrical conductivity of Al-bearing $SiO_2$ with 1750 ppm $H_2O$ at 33–36 GPa, measured using platinum foil electrodes without gold sputtering. Apart from the electrode materials, all other conditions, such as the cell assemblies used for the electrical conductivity measurements, were identical to those in other conductivity experiments conducted in this study. The electrical conductivity of the sample initially increased linearly with decreasing reciprocal temperature; however, it plateaued with further temperature increase. The conductivity measured during the cooling path was approximately an order of magnitude lower than that obtained during the heating path. This discrepancy is likely attributed to the gradual dehydration of the sample during heating, which reduces the number of charge carriers and, consequently, the conductivity. The higher activation enthalpy, indicated by the steeper slope in the Arrhenius plot, further supports this conclusion, as activation enthalpy negatively correlates with the water content of the sample (Yoshino, 2010).

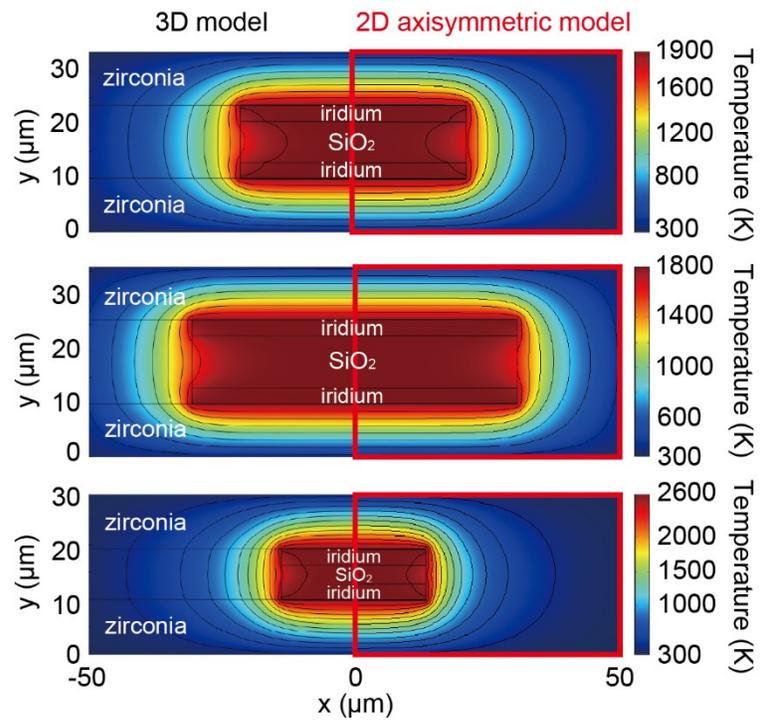

**Figure S2.** Comparison of simulated temperature distributions obtained using the 3D model (left) and the 2D axisymmetric model (right, highlighted by red rectangles). The results demonstrate that the temperature structures derived from both models are consistent.

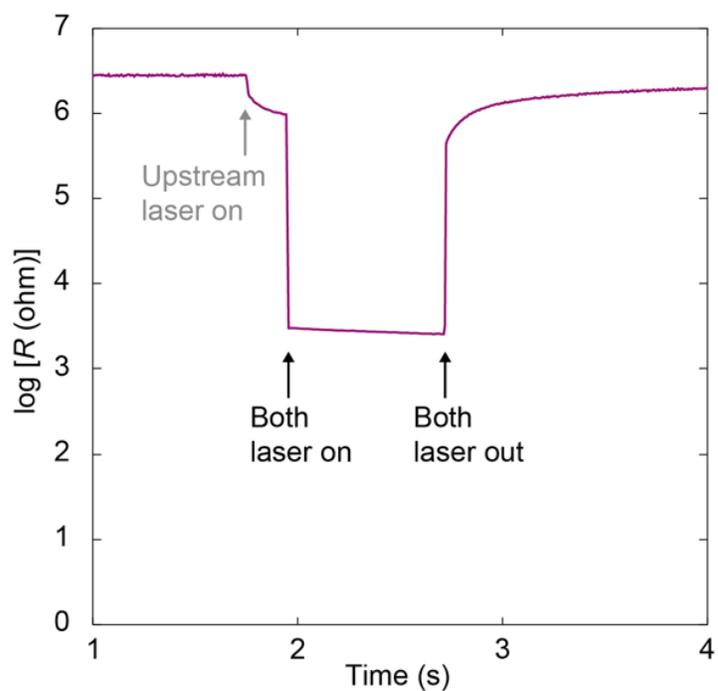

**Figure S3.** Time evolution of sample resistance upon heating to 2610 K at 82 GPa in run #3. Small reduction in resistance just before the significant drop, indicated by the gray arrow, was caused since the laser heating system output the upstream laser ~200 ms earlier than the downstream laser.

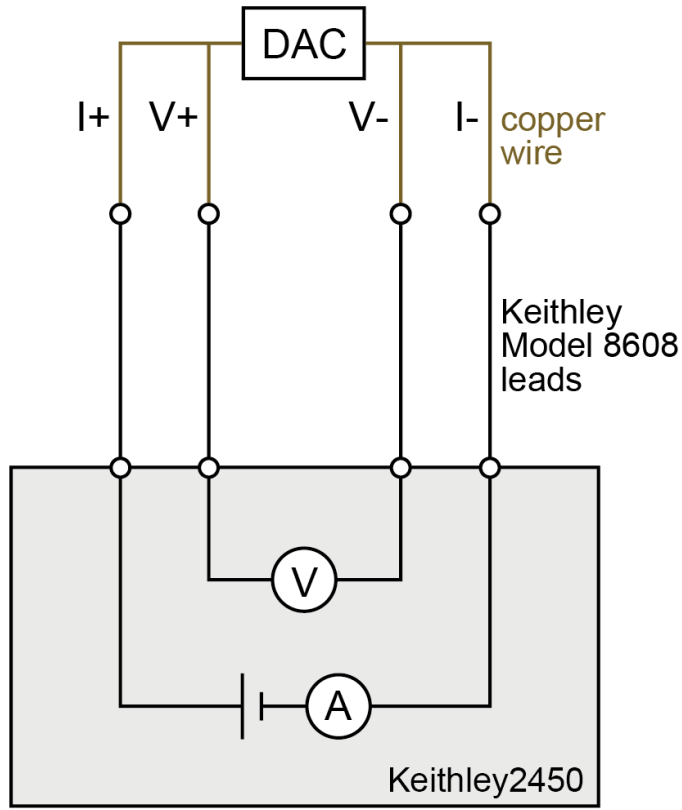

**Figure S4.** Schematic illustration of the pseudo-four-terminal assembly.

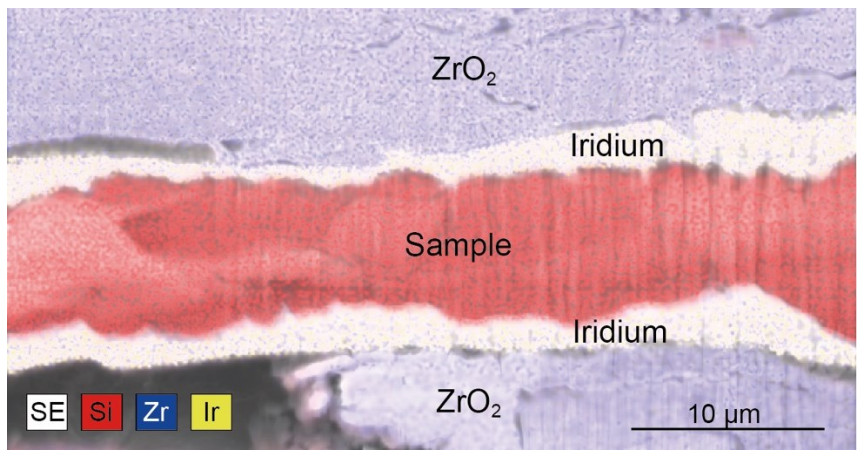

**Figure S5.** Cross section of a sample recovered from run #2 at ~54 GPa. The secondary electron (SE) image was combined with the EDS X-ray maps for Si (red), Zr (blue) and Ir (yellow). Note that these images were acquired from an oblique direction of 52 degrees to the sample stage, and thus the apparent thickness shown here is different from the actual sample thickness given in Table S1.

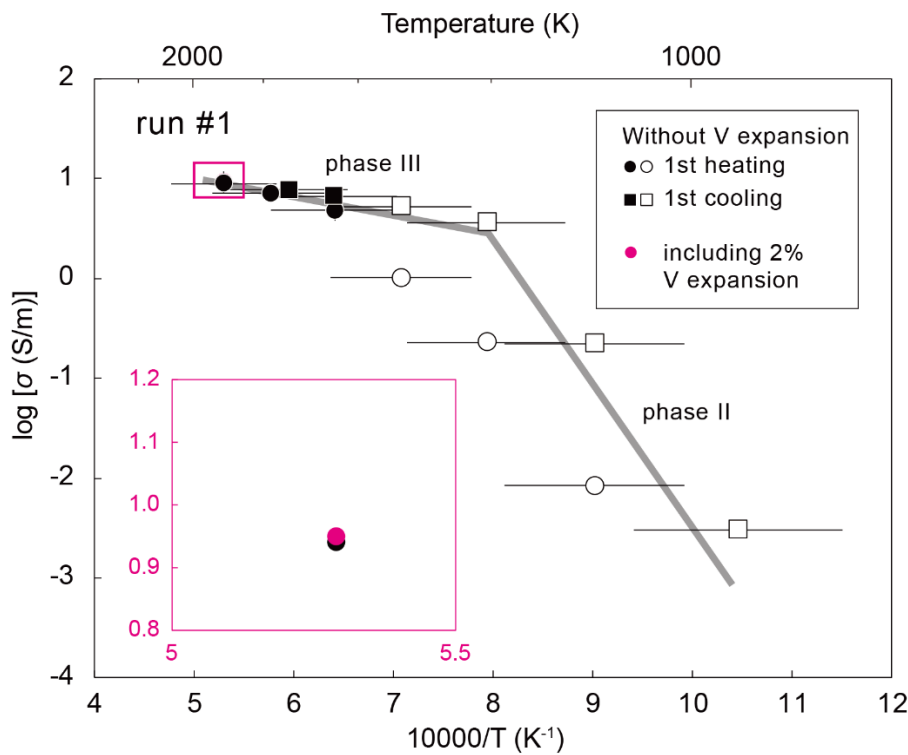

**Figure S6.** Effect of volume expansion on the electrical conductivity estimated in run #1 of this study. The inset shows a zoomed-in view at a reciprocal temperature range of 5 to 5.5 and an electrical conductivity range of $10^{0.8}$ to $10^{1.2}$ (S/m), shown as a magenta rectangle in the figure. The thermal expansion of the sample in run #1 was estimated using the thermal equation of state for stishovite (Fischer et al., 2018). At 46 GPa and 1890 K, the volume of stishovite increases by 2% (the maximum temperature in run #1). Assuming isotropic deformation, the deformation rate in each of the three spatial dimensions (x,y,z) is approximately 0.66%. However, under the extreme assumption that the entire volume expansion occurs solely along the compressional axis in a DAC, the sample thickness would expand by 2%. In this extreme scenario, the electrical conductivity at 46 GPa and 1890 K would increase from 8.73 S/m (black symbol in the inset figure) to 8.91 S/m (magenta symbol in the inset figure). This change is negligible and difficult to distinguish in an Arrhenius plot. Since this estimate is based on an extreme case, the actual effect is expected to be smaller.

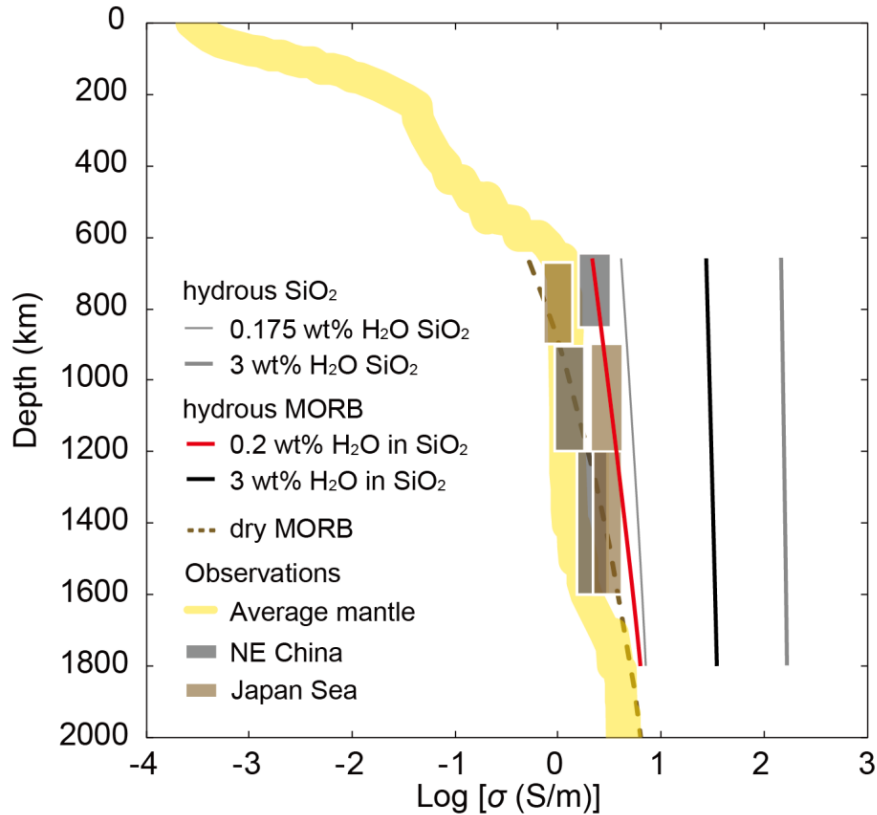

**Figure S7.** Electrical conductivity of hydrous Al-bearing $SiO_2$ and MORB in subducted slab and the comparison with observations. Gray thin and thick lines show the electrical conductivity of superionic Al-bearing $SiO_2$ with 0.175 wt% and 3 wt% $H_2O$, respectively. Red and black bold lines indicate the calculated electrical conductivity of hydrous MORB, including the Al-bearing $SiO_2$ phase with 0.2 wt% and 3 wt% $H_2O$, respectively. Brown broken curve indicates the electrical conductivity for the dry MORB. Yellow band represents the globally-averaged, one-dimensional mantle electrical conductivity model based on geomagnetic observations (Velímský & Knopp, 2021). The gray and brown rectangles show the local observations of high electrical conductivity anomaly underneath NE China at 650–850 km depth (Kelbert et al., 2009; Kuvshinov, 2012; Shimizu et al., 2010) and Japan Sea at 900–1200 km depth (Kelbert et al., 2009; Kuvshinov, 2012), respectively.

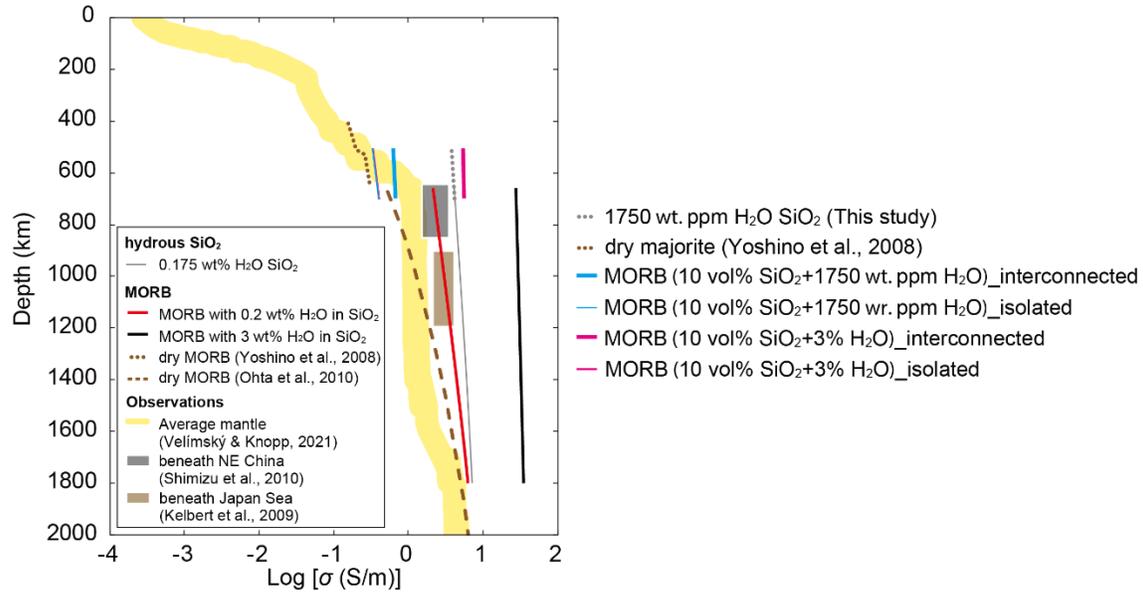

**Figure S8.** Electrical conductivity of dry and hydrous MORB comprising majorite+10 vol% Al-bearing $SiO_2$ stishovite in the mantle transition zone and the uppermost lower mantle. Blue thick line represents the electrical conductivity of majoritic MORB with Al-bearing $SiO_2$ containing 1750 wt ppm $H_2O$ under the assumption of an interconnected model, while the blue thin line corresponds to the non-interconnected model. Pink thick line represents the electrical conductivity of majoritic MORB with Al-bearing $SiO_2$ containing 3 wt% $H_2O$ under the assumption of an interconnected model, while the pink thin line corresponds to the non-interconnected model. Gray thin and thick lines show the electrical conductivity of superionic Al-bearing $SiO_2$ with 0.175 wt% and 3 wt% $H_2O$, respectively. Red and black bold lines indicate the calculated electrical conductivity of hydrous MORB, including the Al-bearing $SiO_2$ phase with 0.2 wt% and 3 wt% $H_2O$, respectively. Brown dotted line represents the electrical conductivity of dry majoritic MORB (Yoshino et al., 2008). Brown dashed line shows the electrical conductivity estimated based on post-majorite MORB data (Ohta et al., 2010). Yellow band represents the globally-averaged, one-dimensional mantle electrical conductivity model based on geomagnetic observations (Velímský & Knopp, 2021). The gray and brown rectangles show the local observations of high electrical conductivity anomaly underneath NE China at 650–850 km depth (Kelbert et al., 2009; Kuvshinov, 2012; Shimizu et al., 2010) and Japan Sea at 900–1200 km depth (Kelbert et al., 2009; Kuvshinov, 2012), respectively.

**Text S1.** The current reversal technique

We applied a specified current of ±1 mA to a sample and measured voltage (Ohta et al., 2023). When applying a current $I_0$ = +1 mA, the following equation holds based on Ohm's rule;

$$I_0 \times R_{sample} + \Delta V_S = V_{meas+} \quad \text{(S1)}$$

where $\Delta V_S$ is Seebeck voltage, and $V_{meas+}$ is measured voltage. Upon current reversal, the direction of the current changes, but a temperature gradient field stays unchanged to the first approximation, and thus Ohm's rule can be written as;

$$-I_0 \times R_{sample} + \Delta V_S = V_{meas-} \quad \text{(S2)}$$

where $V_{meas-}$ is the voltage measured upon current reversal. By removing $\Delta V_S$ from Eqs. S1 and S2,

$$2 \times I_0 \times R_{sample} = V_{meas+} - V_{meas-} \quad \text{(S3)}$$

$$R_{sample} = (V_{meas+} - V_{meas-}) / (2 \times I_0) \quad \text{(S4)}$$

Therefore, the effect of Seebeck voltage can be cancelled out by dividing the difference between the two readings of the measured resistance ($V_{meas+}/I_0$ - $V_{meas-}/I_0$) by two under the current reversal.

**Text S2.** Extraction of the conductivity anomaly value beneath NE China from a color map

The color map conductivity data in Shimizu et al. was given as a relative conductivity difference of $\sigma/\sigma_{\text{1-D average (1-D NP2J1D model)}}$. We carefully extracted the relative conductivity data based on the color mapping results using our color-matching Python code, in the anomaly region (50 pixels×50 pixels red rectangle region in Figure S9) by computing its two-dimensional average.

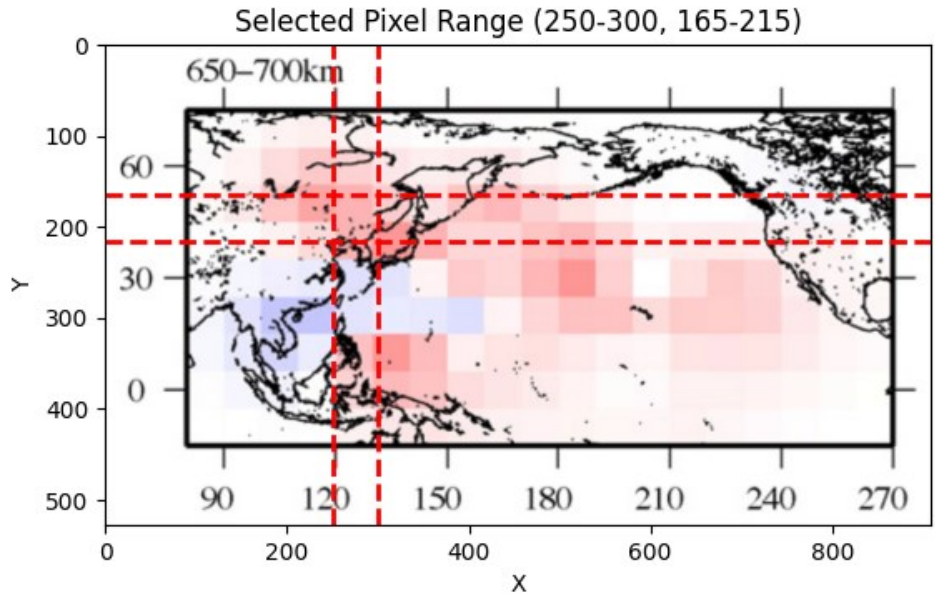

**Figure S9.** Selected pixel area to estimate the relative electrical conductivity beneath NE China. The color mapping results were from Shimizu et al. (2010).

To obtain the absolute conductivity values beneath NE China from Shimizu et al.'s data, we used the extracted relative conductivity value and the average conductivity from the 1-D conductivity model (NP2J1D) in Shimizu et al. (2010). We checked that the difference in conductivity models between NP2J1D (Shimizu et al., 2010) and the average mantle model illustrated in Figure 8a in the main manuscript (Velímský, J. & Knopp, 2021) was similar in the depths deeper than 400 km (see yellow band and orange dotted line in Figure S10). The black rectangles in Figure S10 shows the obtained conductivities beneath NE China from Shimizu et al. (2010)'s color map conductivity data. The error bars indicate the 1σ uncertainty in the 50 pixels×50 pixels region.

We also obtained the conductivity values beneath the Japan Sea using the same method from color map data reported in Kelbert et al. (2009).

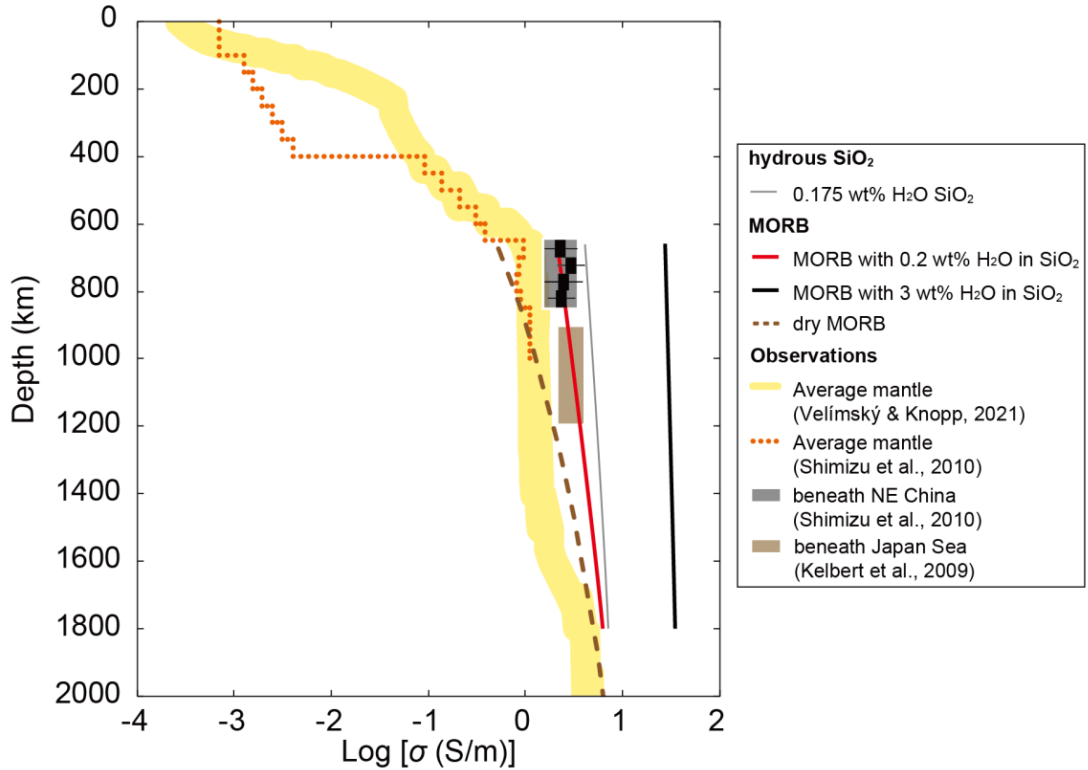

**Figure S10.** Electrical conductivity of hydrous Al-bearing $SiO_2$ and MORB in subducted slab and the comparison with observations. Gray thin and thick lines show the electrical conductivity of superionic Al-bearing $SiO_2$ with 0.175 wt% and 3 wt% $H_2O$, respectively. Red and black bold lines indicate the calculated electrical conductivity of hydrous MORB, including the Al-bearing $SiO_2$ phase with 0.2 wt% and 3 wt% $H_2O$, respectively. Brown broken curve indicates the electrical conductivity for the dry MORB. Yellow band and orange dotted line represent the globally-averaged, one-dimensional mantle electrical conductivity model based on geomagnetic observations reported in Velímský & Knopp (2021) and Shimizu et al. (2010), respectively. The gray and brown rectangles show the local observations of high electrical conductivity anomaly underneath NE China at 650–850 km depth (Kelbert et al., 2009; Kuvshinov, 2012; Shimizu et al., 2010) and Japan Sea at 900–1200 km depth (Kelbert et al., 2009; Kuvshinov, 2012), respectively. The black rectangles represent the conductivities beneath NE China extracted from Shimizu et al. (2010).

**Table S1. Measured and simulated temperatures in this study.**

| Run # | Sample thickness (µm) | Laser spot size (µm) | Measured *T* (K) | Simulated sample *T* (K) |
|---|---|---|---|---|
| 1 | 7.67(87) | 21.3(21) | 1130(110) | 1110(110) |
|   |          |          | 1280(130) | 1260(130) |
|   |          |          | 1440(140) | 1410(140) |
|   |          |          | 1590(160) | 1560(160) |
|   |          |          | 1770(180) | 1740(180) |
|   |          |          | 1920(190) | 1890(190) |
|   |          |          | 1710(170) | 1680(170) |
|   |          |          | 1590(160) | 1570(160) |
|   |          |          | 1440(140) | 1410(140) |
|   |          |          | 1280(130) | 1260(130) |
|   |          |          | 1130(110) | 1110(110) |
|   |          |          | 1130(110) | 1110(110) |
|   |          |          | 970(100)  | 960(100)  |
| 2 | 9.90(130) | 30.6(31) | 1210(120) | 1190(120) |
|   |          |          | 1220(120) | 1210(120) |
|   |          |          | 1280(130) | 1260(130) |
|   |          |          | 1410(140) | 1390(140) |
|   |          |          | 1800(180) | 1780(180) |
|   |          |          | 1820(180) | 1800(180) |
|   |          |          | 1380(140) | 1360(140) |
|   |          |          | 1260(130) | 1240(130) |
|   |          |          | 1190(120) | 1170(120) |
|   |          |          | 1080(110) | 1070(110) |
|   |          |          | 1040(100) | 1030(100) |
|   |          |          | 960(100)  | 950(100)  |
|   |          |          | 880(90)   | 870(90)   |
|   |          |          | 800(80)   | 790(80)   |
|   |          |          | 930(90)   | 920(90)   |
|   |          |          | 1070(110) | 1050(110) |
|   |          |          | 1150(120) | 1130(120) |
|   |          |          | 1280(130) | 1260(130) |
|   |          |          | 1380(140) | 1370(140) |

| Run # | Sample thickness (µm) | Laser spot size (µm) | Measured *T* (K) | Simulated sample *T* (K) |
|---|---|---|---|---|
| 3 | 3.30(80) | 13.3(13) | 1780(180) | 1750(180) |
|   |          |          | 1300(130) | 1280(130) |
|   |          |          | 1270(130) | 1260(130) |
|   |          |          | 1160(120) | 1140(120) |
|   |          |          | 1070(110) | 1050(110) |
|   |          |          | 990(100)  | 970(100)  |
|   |          |          | 910(90)   | 900(90)   |
|   |          |          | 830(80)   | 820(80)   |
|   |          |          | 750(70)   | 740(70)   |
|   |          |          | 670(70)   | 660(70)   |
|   |          |          | 2630(260) | 2610(260) |
|   |          |          | 2420(240) | 2400(240) |
|   |          |          | 2210(220) | 2190(220) |
|   |          |          | 2010(200) | 1990(200) |
|   |          |          | 1800(180) | 1780(180) |
|   |          |          | 1590(160) | 1570(160) |
|   |          |          | 1380(140) | 1370(140) |
|   |          |          | 1170(120) | 1160(120) |
|   |          |          | 960(100)  | 950(100)  |
|   |          |          | 1170(120) | 1160(120) |
|   |          |          | 1380(140) | 1370(140) |
|   |          |          | 1590(160) | 1570(160) |
|   |          |          | 1800(180) | 1780(180) |
|   |          |          | 2010(200) | 1990(200) |
|   |          |          | 2210(220) | 2190(220) |
|   |          |          | 2420(240) | 2400(240) |
|   |          |          | 1400(140) | 1390(140) |
|   |          |          | 1550(160) | 1540(160) |
|   |          |          | 1630(160) | 1610(160) |
|   |          |          | 1780(180) | 1760(180) |
|   |          |          | 2010(200) | 1990(200) |
|   |          |          | 2210(220) | 2190(220) |

**Table S2. Activation enthalpy (eV) of hydrous Al-bearing SiO$_2$.**

| Run # | Pressure (GPa) | Phase I | Phase II | Phase III |
|---|---|---|---|---|
| 1 | 41-46 | — | 2.11(66) | 0.35(10) |
| 2 | 52-57 | 0.67(12) | 1.37(37) | 0.08(4) |
| 3 | 73-82 | 0.46(15) | 1.36(37) | 0.31(88) |

**Table S3. Input laser power in the experiments and COMSOL simulation.**

| Run # | Experiment (W) | COMSOL simulation (W) | Efficiency (%) |
|---|---|---|---|
| 1 | 35 | 13.1 | 37 |
| 2 | 48 | 16.4 | 34 |
| 3 | 74 | 27.5 | 37 |

The discrepancy in input power can be attributed to multiple reflections occurring within the optical system, at the diamond anvil surface, and at the iridium surface in the DAC. The efficiency is calculated as (Power in simulation/Power in experiment)×100%.